# Dielectric function of $(Zn_xCd_{1-x})_3P_2$ alloy system in the region of direct optical transitions

**D.M. Stepanchikov[1], S.V. Shutov[2]**

[1]*Kherson national technical university, department of general and applied physics, research laboratory of the theory of solids, 24, Berislavske shosse, 73008 Kherson, Ukraine, e-mail: step_75@mail.ru*

[2]*V.Lashkaryov Institute of Semiconductor Physics, NAS of Ukraine, 41, prospect Nauky, 03028 Kyiv, Ukraine*

The band structure of $(Zn_xCd_{1-x})_3P_2$ alloy system is considered within the framework of Kildal's band model. Frequency dependencies of real and imaginary parts of dielectric function are received and analyzed in terms of direct band to band transitions. Theoretical calculations were performed for light polarized both parallel and perpendicular to the *c*- axis of the crystal. In calculations the selection rules for optical transitions were applied. The frequency dependence of real part of dielectric function is described by a maximum in $h\nu \approx (1{,}2 \div 1{,}5)E_g$ energy region. In high energy region $h\nu \gg E_g$ the imaginary part of dielectric function has a plateau. Longitudinal dielectric function is less than the transverse dielectric function for all compositions *x* of $(Zn_xCd_{1-x})_3P_2$ alloy system both for real and for imaginary parts. When turning from $Zn_3P_2$ to $Cd_3P_2$ the reduction of dielectric function values is occurs.

**I. Introduction**

$Zn_3P_2$ and $Cd_3P_2$ form a continuous series of substitution $(Zn_xCd_{1-x})_3P_2$ solid solutions, which possess to the tetragonal unit cell described by $P4_2/nmc$ symmetry (space symmetry group $D_{4h}^{15}$) and belong to the group of $A_3^{II}B_2^{V}$ compounds. The $(Zn_xCd_{1-x})_3P_2$ solid solutions are characterized by a direct fundamental energy gap in the range from 0,53 eV (for $Cd_3P_2$) to 1,51 eV (for $Zn_3P_2$) with 300 K [1,2]. Therefore, they are drawing particularly strong attention as materials, which will make it possible to produce highly efficient solar cells, sensors, IR lasers, energy converters, ultrasonic multipliers, Li-ion batteries and the like at low cost [3-9].

Possibility of producing of the simple and reliable blue light emitting diodes, electromagnetic radiation amplifiers in optical fiber communication systems, devices with use nonlinear optical effects in the region of absorption edges are perspective practical applications of $Cd_3P_2$ nanocrystals. Nanoparticles of $Cd_3P_2$ which are estimated to be ~1,5-4 nm in diameter, are characterized by a very strong quantum size effects. It is due to the large excitonic diameter of $Cd_3P_2$ (360 Å) in comparison with GaAs (233 Å), InP (216 Å), CdSe (60 Å), CdS (47 Å), and therefore fluorescence of $Cd_3P_2$ nanoparticles are stronger [10,11].

Zinc phosphide is used at direct fabrication of hierarchical nanotube/nanowire heterostructures with controlled morphologies, crystallography, and surface architectures. The hierarchical $Zn_3P_2/ZnS$ one-dimensional nanotube/nanowire heterostructure may be considered as example [12]. $Zn_3P_2$ nanotubes with outer diameters of 100÷200 nm and wall thickness of ~10, 20 and 45 nm, show emissions centered at about 491, 711 and 796 nm respectively [13]. Optoelectronic devices fabricated using single crystalline branched zinc phosphide nanowires demonstrate a high sensitivity and rapid response to impinging light and it offers a great potential for a high efficient spatial resolved photon detector [14]. Study and practical application of the bicrystalline $Zn_3P_2$ and $Cd_3P_2$ nanobelts, which were synthesized at recent time, is a new perspective way [34].

However, compared with the significant progress in bulk and low-dimensional II-VI and III-V semiconductors, research on II-V semiconductors has been lingering far behind due to the lack of appropriate and generalized synthetic methodologies. Now this problem of producing II-V compound semiconductors (including nanoscales) has been substantially decided. Therefore, theoretical studies of the features of band structure of II-V materials once again become an urgent problem. The considering a new characteristics of semiconductors such as size effects, layered crystalline structure, extreme anisotropy, exciton effects are requisite because of fast progress of nanotechnologies [15]. The one-ax symmetry and layered crystalline structure of II-V semiconductors both provide for high anisotropy of physical properties. So, knowledge of behavior of dielectric function in the region of the direct interband transitions and its anisotropy (polarization dependences) are very important. Moreover, the like studies for $Cd_3P_2$ as experimental so and theoretical both for author are not known. The investigations of the IR reflection spectra of $Zn_3P_2$ single crystals in $e_p \perp c$ polarization conditions are carried out and value of $\varepsilon_{\infty\perp}$ is received in [16]. Measurements of reflectivity, absorption and refraction spectra of $Zn_3P_2$ on polarized radiation were executed in [17,18]. Studies of dielectric function of $Cd_3P_2$ in IR region [19,20] and the other optical parameters for $Cd_3P_2$ and $Zn_3P_2$ [2,3,21] were executed on unpolarized light only. While for intermediate compositions of $(Zn_xCd_{1-x})_3P_2$ alloy systems the polarized studies of optical parameters are not known. Absorptivity of $(Zn_xCd_{1-x})_3P_3$ solid solutions with x=0.66, 0.75, 0.91, and 1.0 has been measured in [22]. And photoluminescence spectra of electron beam evaporated $(Zn_xCd_{1-x})_3P_3$ thin films with x=0, 0.2, 0.4, 0.5, and 1 has been recorded in [2].

Dielectric function is connected with other important parameters of the semiconductor such as absorption coefficient, reflection coefficient, refractive index, conductivity. Besides, the dielectric function values powerfully influences upon values of the exciton binding energy. In low dimensionality systems the exciton effects are very strong and appear to be of interest both from practical and theoretical viewpoints. So, the high temperature displaying of exciton effects is not eliminated even into bulk $Zn_3P_2$ crystals [15]. Therefore, the knowledge of polarization dependences of dielectric function in the optical transitions region is very needed for correct calculation of exciton binding energy. Since such studies practically are absent, study of this problem is actual and necessary. This paper presents a theoretical study of the polarization dependencies and the frequency dependencies of dielectric function of $(Zn_xCd_{1-x})_3P_2$ alloy systems with different values $x$ in the region of direct optical transitions. The generalized Kildal's band model [15,23] is used for calculations. The received results are analyzed with famous experimental data.

**II. Band structure**

The effective Hamiltonian for $D_{4h}^{15}$ space modifications of $A_3^{II}B_2^{V}$ semiconductors nearby the point $\Gamma$ and within quasicubic approximation of Kildal's band model may be written as [23]:

$$H_k = \begin{pmatrix} H_{11} & H_{12} \\ H_{12}^+ & H_{22} \end{pmatrix} \qquad (1)$$

here

$$H_{11} = \begin{pmatrix} E_g & 0 & i\eta^{-2}Pk_z & 0 \\ 0 & E_g & 0 & i\eta^{-2}Pk_z \\ -i\eta^{-2}Pk_z & 0 & -\delta - \Delta/3 & 0 \\ 0 & -i\eta^{-2}Pk_z & 0 & -\delta - \Delta/3 \end{pmatrix};$$

$$H_{22} = \begin{pmatrix} -2\Delta/3 & 0 & 0 & 0 \\ 0 & -2\Delta/3 & 0 & 0 \\ 0 & 0 & 0 & 0 \\ 0 & 0 & 0 & 0 \end{pmatrix}; \quad H_{12} = \begin{pmatrix} 0 & iPk_- & 0 & iPk_+ \\ -iPk_+ & 0 & iPk_- & 0 \\ \sqrt{2}\Delta/3 & 0 & 0 & 0 \\ 0 & \sqrt{2}\Delta/3 & 0 & 0 \end{pmatrix} \qquad (2)$$

Thereto: ($E_g$, $\Delta$, $P$) – are three well-known Kane's parameters, i.e. the energy gap, the spin-splitting parameter and the matrix element of impulse respectively; $\delta$ – is the parameter of the crystal field; $\eta$ – is the scalar factor taking into account the tetragonal deformation of the lattice; $k_{\pm} = (1/\sqrt{2})(k_x \pm ik_y)$.

By diagonalization of Hamiltonian (1) we get a following secular equation:
$$(k_x^2 + k_y^2)P^2 f_1(E) + k_z^2 P^2 f_2(E) - \Gamma(E) = 0 \tag{3}$$
where energy polynomials $\Gamma(E)$, $f_1(E)$, $f_2(E)$ are given by relations:
$$\Gamma(E) = E\left[(E - E_g)\left(\left(E + \frac{2\Delta}{3}\right)\left(E + \delta + \frac{\Delta}{3}\right) - \frac{2\Delta^2}{9\eta^2}\right)\right] \tag{4}$$
$$f_1(E) = \left(E + \frac{\Delta}{3}\right)\left(E + \delta + \frac{\Delta}{3}\right) - \frac{\Delta^2}{9\eta^2} \tag{5}$$
$$f_2(E) = E\left(E + \frac{2\Delta}{3}\right)\eta^{-4} \tag{6}$$

Dispersion equation (3) has four non-identical solutions, each with double spin degeneration. These solutions are describing the conductivity band (c), the heavy holes band (hh), the light holes (lh) and the spin-orbital split bands (so), respectively (see figure 1). In addition, the equation (3) is a particular case of second order equation, describing a surface of the rotation around polar $z$- axis in the $k$- space. Therefore, transversal and longitudinal effective masses associate themselves with two semi-axes of this surface [15,24]:
$$m_{\perp}^* = \frac{\hbar^2 \Gamma(E)}{2(E - E_0)P^2 f_1(E)} = \frac{E\hbar^2(E - E_g)((3(E + \delta) + \Delta)(3E + 2\Delta)\eta^2 - 2\Delta^2)}{2P^2(E - E_0)((3E + \Delta)(3(E + \delta) + \Delta)\eta^2 - \Delta^2)} \tag{7}$$
$$m_{\|}^* = \frac{\hbar^2 \Gamma(E)}{2(E - E_0)P^2 f_2(E)} = \frac{\hbar^2 \eta^2 (E - E_g)((3(E + \delta) + \Delta)(3E + 2\Delta)\eta^2 - 2\Delta^2)}{6P^2(E - E_0)(3E + 2\Delta)} \tag{8}$$
where $E_0$ – is the energy of bands extremum in $\Gamma$ ($k = 0$) point [15]
$$E_0^c = E_g;\ E_0^{hh} = 0;\ E_0^{lh,so} = -\frac{3\eta(\delta + \Delta) \mp \sqrt{9\delta^2 \eta^2 - 6\delta\Delta\eta^2 + \Delta^2(8 + \eta^2)}}{6\eta} \tag{9}$$
In the last equation the sign "–" and "+" correspond to lh- band and so- band, respectively. The top of the hh-band is selected as the zero energy coordinates.

The band parameters value depending on composition $x$ of $(Zn_xCd_{1-x})_3P_2$ alloy systems are shown in table 1. Regrettably, in literary sources the values of full set of band parameters are available for $Zn_3P_2$ and $Cd_3P_2$ only. Using values of band gap for $Cd_3P_2$ [25], $Zn_3P_2$ [17] and $(Zn_xCd_{1-x})_3P_2$ with $x$ = 0,2; 0,4; 0,5; 0,8 [2] we have got a following exponential equation for function of $E_g(x)$:
$$E_g(x) = 0{,}53 + 1{,}25x - 1{,}46x^2 + 2{,}21x^3 + 0{,}82x^4 - 1{,}84x^5 \tag{10}$$
Scalar factor $\eta = c/(a\sqrt{2})$ ($a,b$ – are the lattice parameters) for $Cd_3P_2$ and $Zn_3P_2$ has nearly equal values and so it was considered as constant for $(Zn_xCd_{1-x})_3P_2$ alloy systems. The composition dependencies for other band parameters i.e. $\Delta$, $P$ and $\delta$ were accepted as linear, similar to following function
$$A(x) = A(Zn_3P_2)x + A(Cd_3P_2)(1 - x) \tag{11}$$

Undoped $Zn_3P_2$ displays p- type conductivity with characteristic charge concentration about $10^{15} \div 10^{16}$ cm$^{-3}$, while at another end of series $Cd_3P_2$ is the degenerated n- type semiconductor with characteristic charge concentration about $10^{17} \div 10^{18}$ cm$^{-3}$. Data about type of conductivity in the intermediate solutions of the $(Zn_xCd_{1-x})_3P_2$ alloy system are not available. Therefore, we suppose that the crossover p- to n- type is in the solutions with x=0.5. Until x>0.5 prevail the p- type conductivity, and with x<0.5 prevail the n- type conductivity. Using such assumptions we have got a following nonlinear equation for composition dependence of Fermi level $\varepsilon_F(x)$:
$$\varepsilon_F(x) = \varepsilon_F(Cd_3P_2) + Bx - Cx^2 \tag{12}$$

where $B = 0.15$ eV, $C = 0.54$ eV. The Fermi level values for different compositions $x$ of $(Zn_xCd_{1-x})_3P_2$ solid solutions are shown in table 1.

Table 1

The main band parameters of $(Zn_xCd_{1-x})_3P_2$ alloy system

| $x$ | $E_g$, eV | $\Delta$, eV | $P$, $10^{-10}$ eV·m | $\delta$, eV | $\eta$ | $\varepsilon_F$, eV[*] |
|---|---|---|---|---|---|---|
| 0 | 0,53 [25] | 0,150 [26] | 7,2 [27] | 0,023 [9] | 0,99 | 0,59 [25] |
| 0,2 | 0,74 [2] | 0,142 | 6,7 | 0,024 | 0,99 | 0,60 |
| 0,4 | 0,94 [2] | 0,134 | 6,2 | 0,025 | 0,99 | 0,56 |
| 0,5 | 1,06 [2] | 0,130 | 5,95 | 0,026 | 0,99 | 0,53 |
| 0,6 | 1,19 | 0,126 | 5,7 | 0,027 | 0,99 | 0,49 |
| 0,8 | 1,46 [2] | 0,118 | 5,2 | 0,029 | 0,99 | 0,36 |
| 1 | 1,51 [17] | 0,110 [18] | 4,7 [28] | 0,03 [18] | 0,99 | 0,2 [29] |

[*]The top of the heavy holes band is selected as the zero counting of Fermi level value.

### III. Selection rules for optical transitions

In the $D_{4h}^{15}$ space group, the states of $X$ and $Y$ symmetry transform according to $\Gamma_5^-$ representations and the states of $Z$ symmetry according to the $\Gamma_2^-$ representation. Owing to the spin-orbit interaction, the $\Gamma_5^-$ state will split into a $\Gamma_7^+$ and $\Gamma_7^-$ whereas the $\Gamma_2^-$ state becomes $\Gamma_7^-$. The lowest conduction band in quasi-cubic approximation $\Gamma_1$ will transform as $\Gamma_6^+$ (fig.1).

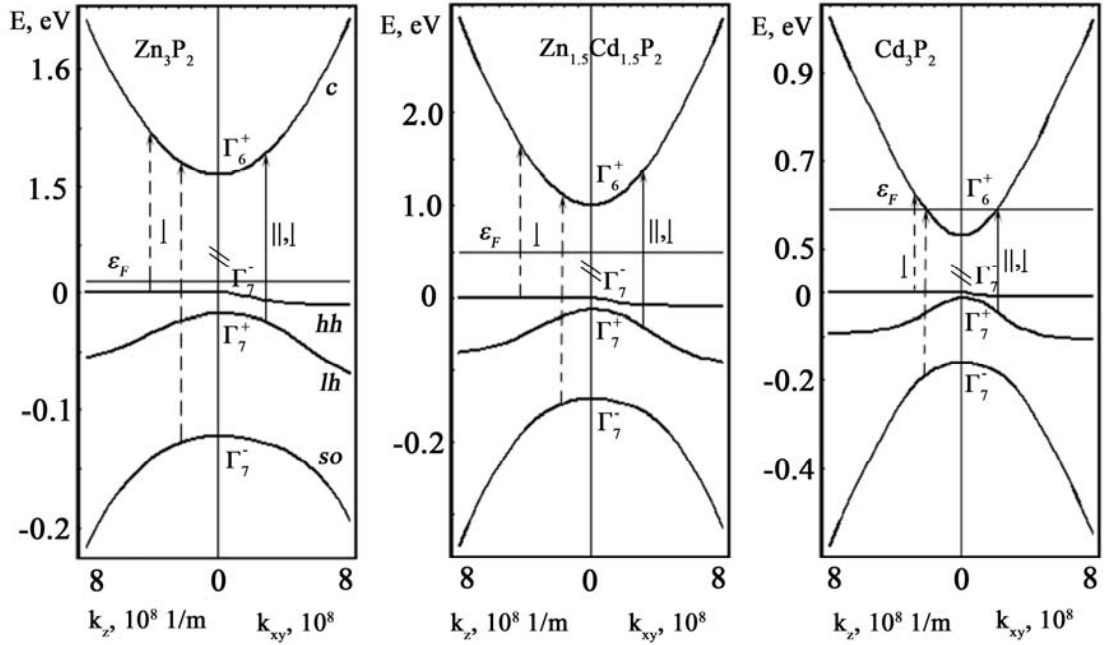

Fig.1. Energy band structure of $(Zn_xCd_{1-x})_3P_2$ alloy system for $x = 1; 0,5; 0$ near the $\Gamma$ point at 300 K. Arrows denotes "allowed" optical transitions for two different light polarization directions.

Knowledge of the selection rules is necessary for interpretations of the optical measurements. In same cases interband transitions can be "forbidden". Furthermore, in the physical system, which has low symmetry, the selection rules can depend on the polarization of light. In the case of the $D_{4h}^{15}$ space group, the perturbation operator for direct optical transitions transforms as $\Gamma_5^-$ for the light polarized perpendicular to the main optical crystalline axis $(e_p \perp c)$ and as $\Gamma_2^-$ for light polarized parallel to the main optical crystalline axis $(e_p \parallel c)$. Thus, the selection rules for $(Zn_xCd_{1-x})_3P_2$ semiconductors depend on the polarization direction of the light. By using the characters of the representation of the elements of $D_{4h}^{15}$ space group, the respective transition rules for "allowed" optical transitions in $\Gamma$ point may be obtained [18]:

$$\begin{aligned} \Gamma_i^\pm \to \Gamma_j^\pm, & \quad for\ e_p \perp c\ and\ e_p \parallel c \\ \Gamma_i^\pm \to \Gamma_j^\mp, & \quad for\ e_p \perp c \end{aligned} \quad (13)$$

So, the optical transition between light holes and conduction bands is "allowed" for all polarization directions of the light, whereas the optical transitions from heavy holes and spin-orbit split bands to conduction band are "allowed" for the light polarized perpendicular $(e_p \perp c)$ to the main crystalline axis only (fig.1). For an "allowed" optical transition the oscillator strength may be determined as [31]

$$F_{cv} = \frac{2|\langle c|ep|v\rangle|^2}{\mu_{cv}(E_c - E_v)} \quad (14)$$

where $|\langle c|ep|v\rangle|^2$ – is the electrical dipole matrix element (probability of interband transition), $\mu_{cv} = m_c^* m_v^* / (m_c^* + m_v^*)$ – is the reduced effective mass ($m_c^*, m_v^*$ – are the effective masses for conduction and valence bands respectively), $E_c, E_v$ – are the energies of conduction and valence bands respectively.

For "forbidden" optical transition the electrical dipole matrix element is a zero in band extreme point only $|\langle c|ep|v\rangle|^2\big|_{k=0} = 0$, but it not mean that the optical absorption does not occurs. In this case is necessary to consider possibility of nonzero values of the electrical dipole matrix element near the $k=0$ point. By distributing the electrical dipole matrix element in Taylor series we get a following expression [31]

$$|\langle c|ep|v\rangle|^2 = \left|\frac{d}{dk}\langle c|ep|v\rangle\right|^2 k^2 + O(k^4) \quad (15)$$

Therefore, in conditions of "forbidden" optical transition the oscillator strength is necessary to consider with electrical dipole matrix element, determined as (15).

The calculated values of oscillator strength for the interband optical transitions in $(Zn_xCd_{1-x})_3P_2$ alloy system are given in table 2.

**IV. Dielectric function**

When the relaxations frequency of the carriers in comparison with photon frequency is negligible quantity, the interband contribution to real part of dielectric function with spherical coordinates $(k, \theta, \varphi)$ has a such form [30,31]

$$\varepsilon_r(\omega) = 1 + \frac{2e^2}{4\pi\varepsilon_0 \pi^2} \sum_j \int_0^\infty \int_0^{2\pi} \int_0^\pi \frac{(f_{vj} - f_c)F_{cvj}|\upsilon_{cvj}|^2 k^2 \sin\theta}{(E_c - E_{vj})((E_c - E_{vj})^2 - (\hbar\omega)^2)} d\theta d\varphi dk \quad (16)$$

The interband contribution to imaginary part of dielectric function in this case is [31,32]

$$\varepsilon_i(\omega) = \frac{\pi e^2 \hbar^2}{2(2\pi)^3 \varepsilon_0} \sum_j \int_0^\infty \int_0^{2\pi} \int_0^\pi \frac{(f_{vj} - f_c)F_{cvj} k^2 \sin\theta}{\mu_{cv}(E_c - E_{vj})} \delta(E_c - E_{vj} - \hbar\omega) d\theta d\varphi dk \quad (17)$$

where $j$ – is the number of valence band, $\varepsilon_0$ – is the electric constant, $e$ – is the elementary charge, $f_{c,v}, \upsilon_{cv}$ – are the Fermi-Dirac function and velocity of charges, which are defined as follows

$$f_{c,v} = \frac{1}{\exp\left(\dfrac{E_{c,v} - \varepsilon_F}{k_0 T}\right)} \quad (18)$$

$$\upsilon_{cv} = \frac{\partial E_c}{\partial k} - \frac{\partial E_v}{\partial k} \qquad (19)$$

The integral on wave vector for real part of dielectric function (16) has a logarithmic break up on top limit. However, the main contribution to integral derives from $E_g \leq E \leq E_{at}$ energy region. So, with logarithmic accuracy the integral must be truncate on $k_m$ value, which corresponds to atomic energy $E_{at}$ [30]. In our calculations there was accepted $E_{at} \approx 4$ eV for $Zn_3P_2$ and $E_{at} \approx 17$ eV for $Cd_3P_2$ [1,35]. The values of wave vector $k_m$, corresponding to $E_{at}$ are presented in table 2. The composition dependency of the wave vector $k_m$ is accepted linear like to equation (11).

Table 2

The interband and dielectric parameters of $(Zn_xCd_{1-x})_3P_2$ alloy system

| $x$ | | $hh \to c$ | $lh \to c$ | $so \to c$ | $k_m$, $10^{10}$ m$^{-1}$ | $\varepsilon_{r\perp}/\varepsilon_{r\parallel}$ **) | $\varepsilon_{i\perp}/\varepsilon_{i\parallel}$ +) |
|---|---|---|---|---|---|---|---|
| 0 | $\mu_\perp$ *) | 0,09 | 0,05 | 0,06 | 2,40 | 11,35 / 7,31 | 1,85 / 1,02 |
| | $\mu_\parallel$ | 0,17 | 0,12 | 0,07 | | | |
| | $F_{cv\perp}$ | 0,39 | 0,35 | 0,26 | | | |
| | $F_{cv\parallel}$ | 1,9·10$^{-3}$ | 0,79 | 8,6·10$^{-2}$ | | | |
| 0,2 | $\mu_\perp$ | 0,14 | 0,08 | 0,09 | 2,06 | 11,76 / 8,23 | 1,98 / 1,12 |
| | $\mu_\parallel$ | 0,25 | 0,18 | 0,11 | | | |
| | $F_{cv\perp}$ | 0,37 | 0,35 | 0,28 | | | |
| | $F_{cv\parallel}$ | 1,2·10$^{-3}$ | 0,74 | 3,7·10$^{-2}$ | | | |
| 0,4 | $\mu_\perp$ | 0,20 | 0,12 | 0,13 | 1,72 | 12,56 / 9,49 | 2,14 / 1,25 |
| | $\mu_\parallel$ | 0,35 | 0,25 | 0,16 | | | |
| | $F_{cv\perp}$ | 0,36 | 0,36 | 0,29 | | | |
| | $F_{cv\parallel}$ | 6,6·10$^{-4}$ | 0,70 | 1,0·10$^{-2}$ | | | |
| 0,5 | $\mu_\perp$ | 0,25 | 0,15 | 0,16 | 1,55 | 13,08 / 10,23 | 2,22 / 1,32 |
| | $\mu_\parallel$ | 0,42 | 0,29 | 0,20 | | | |
| | $F_{cv\perp}$ | 0,36 | 0,33 | 0,29 | | | |
| | $F_{cv\parallel}$ | 4,4·10$^{-4}$ | 0,62 | 3,7·10$^{-3}$ | | | |
| 0,6 | $\mu_\perp$ | 0,30 | 0,19 | 0,20 | 1,38 | 13,64 / 11,08 | 2,31 / 1,39 |
| | $\mu_\parallel$ | 0,50 | 0,35 | 0,25 | | | |
| | $F_{cv\perp}$ | 0,36 | 0,32 | 0,30 | | | |
| | $F_{cv\parallel}$ | 2,8·10$^{-4}$ | 0,59 | 8,9·10$^{-4}$ | | | |
| 0,8 | $\mu_\perp$ | 0,43 | 0,27 | 0,28 | 1,04 | 13,75 / 11,73 | 2,52 / 1,60 |
| | $\mu_\parallel$ | 0,72 | 0,49 | 0,37 | | | |
| | $F_{cv\perp}$ | 0,35 | 0,38 | 0,31 | | | |
| | $F_{cv\parallel}$ | 1,1·10$^{-4}$ | 0,77 | 2,7·10$^{-5}$ | | | |
| 1 | $\mu_\perp$ | 0,55 | 0,35 | 0,36 | 0,70 | 13,82 / 12,06 | 2,76 / 1,82 |
| | $\mu_\parallel$ | 0,91 | 0,60 | 0,49 | | | |
| | $F_{cv\perp}$ | 0,35 | 0,40 | 0,33 | | | |
| | $F_{cv\parallel}$ | 6,9·10$^{-5}$ | 0,85 | 1,5·10$^{-5}$ | | | |

*) The values of the effective masses are expressed in free electron mass;
**) The maximum value of real part of dielectric function;
+) The value of imaginary part of dielectric function on a plateau at $hv \gg E_g$.

With axial symmetry and characteristics of δ-function the equations (16,17) may be a simplified to following forms

$$\varepsilon_r(\omega) = 1 + \frac{4e^2}{\varepsilon_0 \pi^2} \sum_j \int_0^{k_m} \int_0^{\pi/2} \frac{(f_{vj} - f_c) F_{cvj} |v_{cvj}|^2 k^2 \sin\theta}{(E_c - E_{vj})((E_c - E_{vj})^2 - (\hbar\omega)^2)} d\theta dk \qquad (20)$$

$$\varepsilon_i(\omega) = \frac{e^2 \hbar^2}{2\pi\varepsilon_0} \sum_j \int_0^{\pi/2} \frac{(f_{vj} - f_c) F_{cvj} k^2 \sin\theta}{\mu_{cv}(E_c - E_{vj})|v_{cvj}|} d\theta \qquad (21)$$

Here the spin degeneration of each band is taken into account (the degeneration factor is 2).

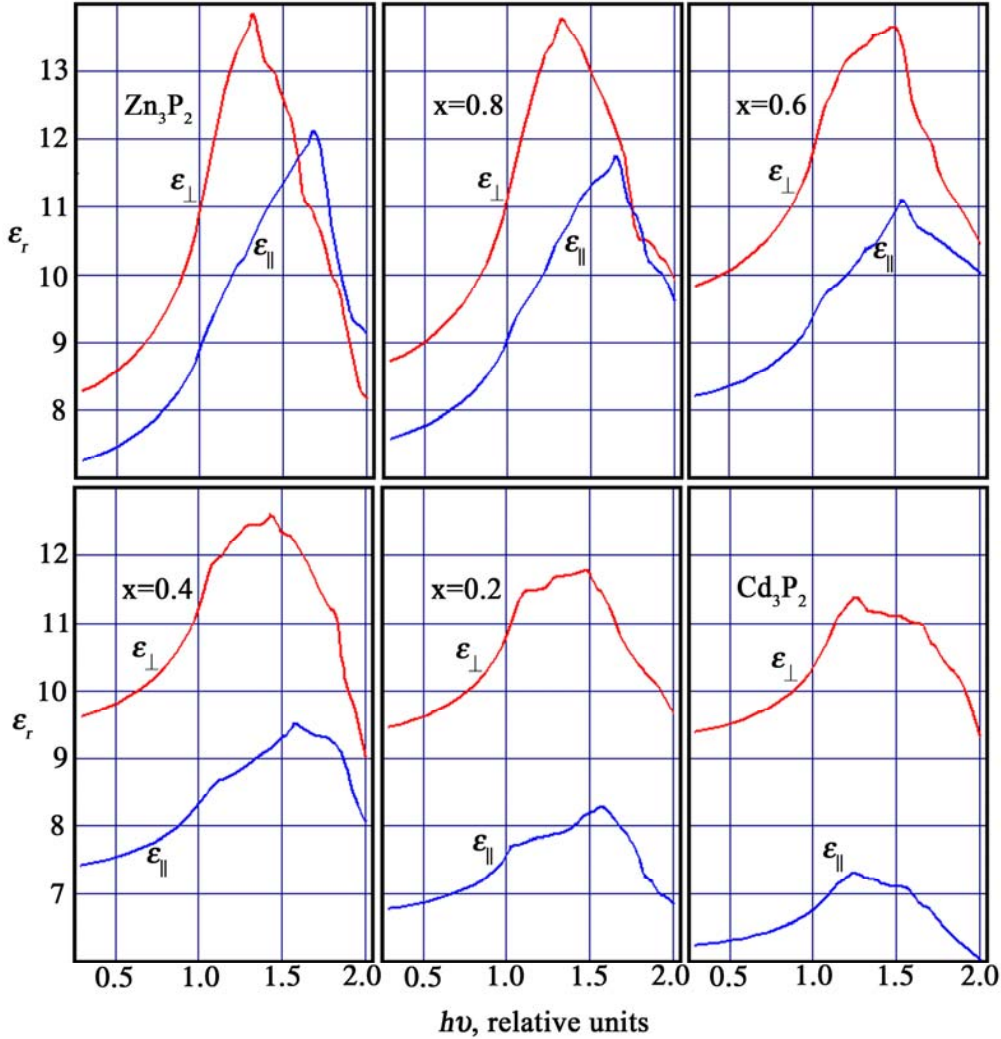

Fig.2. The real part of dielectric function in region of optical transitions for $(Zn_xCd_{1-x})_3P_2$ alloy system vs photon energy (in units of the band gap $E_g$) at $T = 300$ K.

## V. Results and discussion

The generalized Kildal's band model allows defining the polarization dependencies of dielectric function. By substituting in equations (14,18-21) the effective masses and band energies along corresponding directions in the crystal we get the longitudinal and transverse dielectric functions of the considering semiconductors. In addition, the intraband transitions were neglected. The results of our calculations for $(Zn_xCd_{1-x})_3P_2$ alloy system are presented in table 2 and on fig.2,3.

The existence of maximum in $hv \approx (1,2 \div 1,5)E_g$ energy region for real part of dielectric function was found (fig.2). When turning from $Zn_3P_2$ to $Cd_3P_2$ the reduction of values of this maximum is occurs (table 2). Longitudinal real dielectric function is less than the transverse real dielectric function for all compositions $x$ of $(Zn_xCd_{1-x})_3P_2$ alloy system.

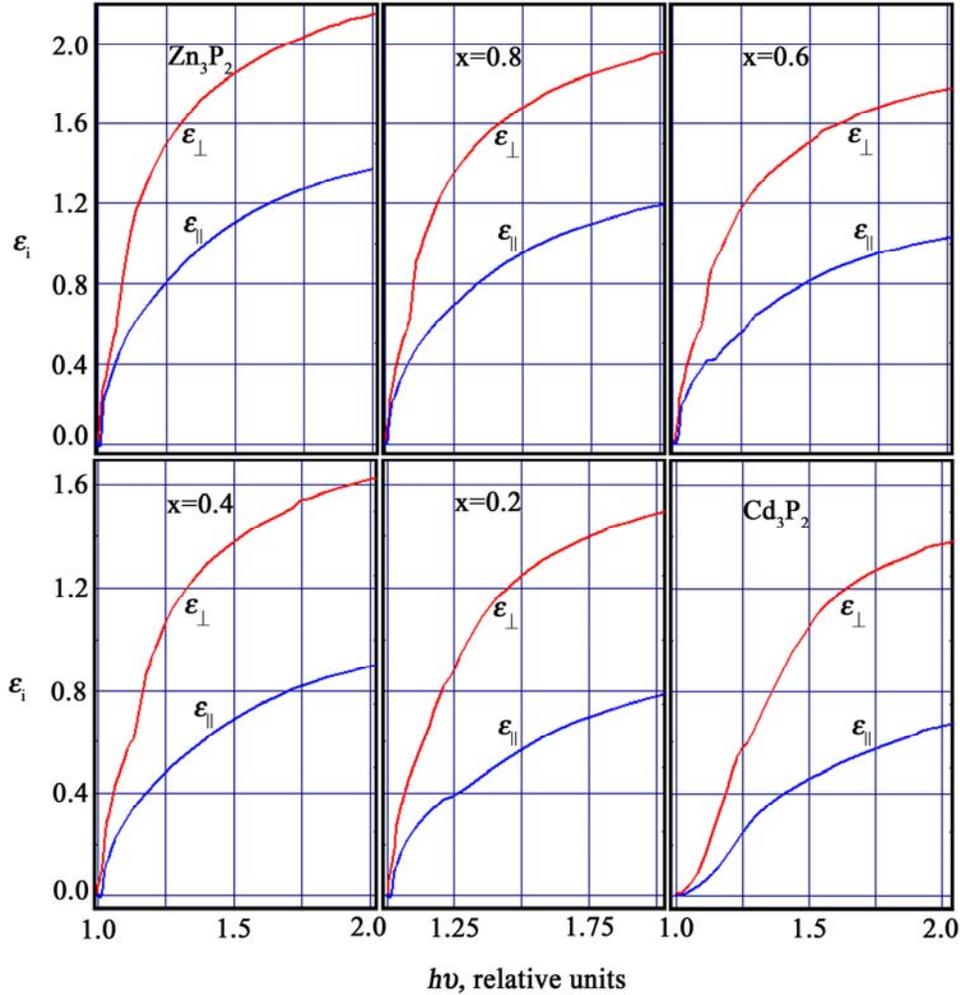

Fig.3. The imaginary part of dielectric function in region of optical transitions for $(Zn_xCd_{1-x})_3P_2$ alloy system vs photon energy (in units of the band gap $E_g$) at $T = 300$ K.

Comparison of available experimental data with our theoretical results testifies to an acceptable correlation between them. From infrared and far-infrared reflectivity measurements are available following values of optical constant $\varepsilon_\infty$: for $Cd_3P_2$ $\varepsilon_\infty = 14$ [20], for $Zn_3P_2$ $\varepsilon_\infty = 10,9$ [21], $\varepsilon_{\infty\perp} = 15,13$ [16]. The measurements of reflectance spectra over a very large range ($0,083 \leq hv \leq 21$ eV) of photon energy are displaying location of the maximum of real part of dielectric function of $Cd_3P_2$ at energy region $\sim 1,5E_g$, and value of maximum about 16 [35]. Calculation of values of the optical constant via refractive index $\varepsilon_\infty = n_\infty^2$ gives following results: for $Cd_3P_2$ $n = 3,6 \Rightarrow \varepsilon_\infty = 12,96$ [25], for $Zn_3P_2$ $n \approx 4,4 \Rightarrow \varepsilon_\infty \approx 19,36$ [36]. It is necessary to note, that the dielectric function $\varepsilon_\infty$ may be considered as the sum of two separate contributions: the electronic contribution, which analyzed in this work and the phonon contribution. In the high frequency region the phonon contribution is proportionate to $\omega_{TO}^2/\omega^2$ (where $\omega_{TO}$ – is the frequency of TO phonon). In our case the photon frequencies $\omega$ are larger than phonon frequencies $\omega_{TO}$. So the phonon contribution to the dielectric function may be neglected in our work.

The monotonous growth at the beginning of interband transitions (fig.3) and the run-out on plateau in high energy region ($hv \gg E_g$) is existing for the imaginary part of dielectric function. The linear dependence of the electronic spectrum of $(Zn_xCd_{1-x})_3P_2$ alloy system in high energy region ($hv \gg E_g$) is the cause of this plateau. Such behavior of imaginary part of the dielectric function is exists for InN and $A_2B_6$ semiconductors for instance [30].

The existence of plateau on frequency dependence of imaginary part of dielectric function is confirmed by experimental data for $Cd_3P_2$: in the energy region $hv \approx (20 \div 30)E_g$ is received $\varepsilon_i \approx 1,3 \div 1,8$ [35]. Calculation of imaginary part of dielectric function via experimental values of refractive index $n$ and extinction coefficient $k$ ($\varepsilon_i = 2nk$) gives following results for $Zn_3P_2$: $\varepsilon_i \approx 2,03$; ($n \approx 4,05$; $k \approx 0,25$; $hv \approx 2$ eV) [36], $\varepsilon_i \approx 1,48$; ($n \approx 3,9$; $k \approx 0,19$; $hv \approx 2,02$ eV) [21].

Absorption coefficient is defined by the dielectric functions as [32]

$$\alpha(\omega) = \frac{\omega}{c}\sqrt{2\left(\sqrt{\varepsilon_i^2(\omega)+\varepsilon_r^2(\omega)} - \varepsilon_r(\omega)\right)} \quad (22)$$

where $c$ – is the light velocity.

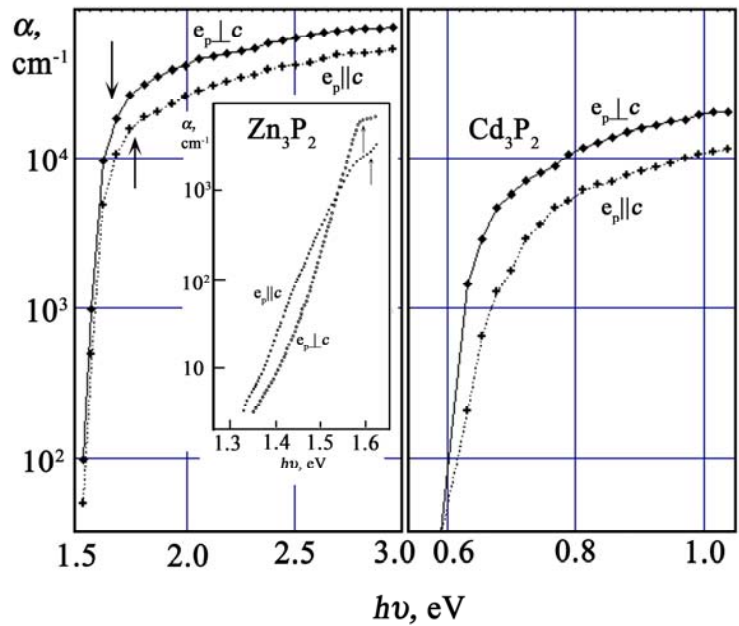

Fig.4. Fundamental absorption edges of $Zn_3P_2$ and $Cd_3P_2$ single crystals calculated for polarized light at $T = 300K$. The insert shows the experimental absorption edge of $Zn_3P_2$ [18]. Arrows denote the regions of good coincidence of experimental and theoretical data.

The stronger values of transverse dielectric function in comparison with longitudinal dielectric function are confirmed by experimental values of absorption coefficient of $Zn_3P_2$. Absorption for $e_p \perp c$ in the range $hv \geq E_g$ is a few times stronger than for $e_p \parallel c$ (see insert on fig.4) [17,18]. For direct interband transitions, one expects no absorption below the energy gap, but in practice on account of the Colomb interaction one usually finds a near-exponentially increasing absorption edge fulfilling the well-known Urbach rule (see $h\nu < E_g$ energy region on insert on fig.4). It was noticed that below the energy gap the experimental absorption for $e_p \perp c$ is less than for $e_p \parallel c$. Regrettably, the like experimental studies for the other members of $(Zn_xCd_{1-x})_3P_2$ alloy system are not known. The calculated absorption spectra near the fundamental edge for $Zn_3P_2$ and $Cd_3P_2$ crystals are presented in fig.4. The Colomb interaction, which gives a broadening of the absorption edge, was neglected in our calculations.

## VI. Conclusions

In summary, we would like to lay down the following conclusions:

1. The dielectric function of $(Zn_xCd_{1-x})_3P_2$ alloy system in optical transitions region has been theoretically established in the framework of Kildal's band model. The possible dependency of band parameters from composition $x$ is offered. In calculations, the selection rules for optical transitions were applied. In addition, the phonon contributions to the dielectric function were neglected. Our theoretical results well correlate with available experimental data.

2. The integral on wave vector for real part of dielectric function (16) has a logarithmic break up on the top limit and therefore must be truncate on $k_m$ value, which corresponds to atomic energy $E_{at}$. Accuracy of the choice of $k_m$ value is a vastly influenced upon final result. Accordingly, the $k_m$ value must be is clearly motivated theoretically; otherwise, it must be used as a fitting parameter.

3. Appreciable anisotropy for the dielectric functions and absorption spectrums within the direct interband optical transitions was found. Longitudinal dielectric function is less than the transverse dielectric function for all compositions $x$ of $(Zn_xCd_{1-x})_3P_2$ alloy system both for real and for imaginary parts. Distinction between longitudinal and transverse dielectric functions is stronger for $Cd_3P_2$.

4. When turning from $Zn_3P_2$ to $Cd_3P_2$ the reduction of dielectric function values at same photon energy is occurs. The frequency dependence of real part of dielectric function is described by a few peaks with maximum in $h\nu \approx (1{,}2 \div 1{,}5)E_g$ energy region. In high energy region $h\nu \gg E_g$ the imaginary part of dielectric function has a plateau. It is caused by linear dependence of the electronic spectrum of $(Zn_xCd_{1-x})_3P_2$ alloy system in high energy region.

5. The calculated absorption coefficient for $Zn_3P_2$ in the energy range $h\nu \geq E_g$ well coincide with experimental data for both light polarization directions. It testifies of rationality of source hypothesis in our theoretical calculations.